# Condensation Calculations in Planetary Science and Cosmochemistry

**Denton S. Ebel**
Department of Earth & Planetary Sciences, American Museum of Natural History
Central Park West at 79th St., New York, NY 10024 USA

Department of Earth & Environmental Sciences, Columbia University
New York, NY 10027 USA

Department of Earth & Environmental Sciences, Graduate Center
City University of New York
New York, NY 10027 USA

## Summary

The Sun's chemical and isotopic composition records the composition of the solar nebula from which the planets are thought to have formed. If a piece of the Sun is cooled to 1000 K at one millibar total pressure, a mineral assemblage is produced that is consistent with those minerals found in the least equilibrated (most chemically heterogeneous), oldest, and compositionally Sun-like (chondritic) hence most "primitive" meteorites. This is an equilibrium or fractional condensation experiment. The result can be simulated by calculations using equations of state for hundreds of gaseous molecules, condensed mineral solids, and silicate liquids, the products of a century of experimental measurements and recent theoretical studies. Such calculations have revolutionized our understanding of the chemistry of the cosmos.

The mid-20th Century realization that meteorites are fossil records of the early Solar System made chemistry central to understanding the origin of the Earth, Moon, and other bodies. Thus "condensation", more generally the distribution of elements and isotopes between vapor and condensed solids and/or liquids at or approaching chemical equilibrium, came to deeply inform discussion of how meteoritic and cometary compositions bear on the origins of atmospheres and oceans and the differences in composition among the planets. This expansion of thinking has had profound effects upon our thinking about the origin and evolution of Earth and the other worlds of our Solar System.

Condensation calculations have also been more broadly applied to protoplanetary disks around young stars, to the mineral "rain" of mineral grains expected to form in cool dwarf star atmospheres, to the expanding circumstellar envelopes of giant stars, to the vapor plumes expected to form in giant planetary impacts, and to the chemically and isotopically distinct "shells" computed and observed to exist in supernovae. The beauty of equilibrium condensation calculations is that the distribution of elements between gaseous molecules, solids and liquids is fixed by temperature, total pressure and the overall elemental composition of the system. As with all sophisticated calculations, there are inherent caveats, subtleties, and computational difficulties.

In particular, local equilibrium chemistry has yet to be consistently integrated into gridded, dynamical astrophysical simulations so that effects like the blocking of light and heat by grains (opacity), absorption and reemission of light by grains (radiative transfer), and buffering of heat by grain evaporation/condensation are fed back into the physics at each node or instance of a gridded calculation over time. A deeper integration of thermochemical computations of chemistry with physical models makes the prospect of a general protoplanetary disk model as hopeful in the 2020s as a general circulation model for global climate may have been in the early 1970's.

## Motivation

What is condensation? In the astronomical context, "condensation" classically describes the sequence of solids and liquids that form (i.e., become thermodynamically stable) with slow, steady cooling of a parcel of vapor of solar bulk composition at constant total pressure ($P^{tot}$). Such conditions were once thought (Cameron, 1962) to characterize the early solar nebula conceptually originated by Kant (1755; cf. Brush, 1999). Condensation calculations for these conditions successfully predict many of the minerals found in the least equilibrated (most primitive) chondritic meteorites (e.g., Grossman, 1972, MacPherson et al., 1984, see figure 1).

However, the picture is more complicated. First, many meteoritic minerals have been affected by chemical processing (i.e., element exchange or equilibration) when their source "parent" bodies were still hot and/or wet. Meteorite petrologists have "seen through" these effects to understand much of the primary mineralogy of the most primitive meteorites. Second, the assumption of a primordial, hot, isotopically and chemically uniform vapor of solar composition resulting from the atomic-scale mixing (e.g., by evaporation) of the molecular cloud material from which the solar accretion disk (the solar nebula) formed (Suess, 1965) faded with the discovery of oxygen isotopic anomalies (Clayton et al., 1973) and noble gas anomalies (Black & Pepin, 1969; Lewis et al., 1975) leading to the discovery of presolar grains preserved in primitive meteorites (Bernatowicz et al., 1987; Lewis et al., 1987). Third, the meteorites least altered on their parent bodies all contain individual components that are far from chemical equilibrium with each other even though they contain suites of minerals consistent with condensation calculations (Ebel et al., 2016).

Nevertheless, the classic condensation sequence - properly calculated - provides a reliable guide to the relative refractory and volatile nature of the elements, with many applications in planetary science and astronomy. Evidence of direct condensation from hot vapor has been credibly claimed in both interplanetary dust containing enstatite $MgSiO_3$ "whiskers" (Bradley et al., 1983; Ogliore et al., 2020), refractory inclusions in meteorites (e.g., Han & Brearley, 2016, 2017; MacPherson et al., 1984; Meibom et al., 2007), and presolar grains (Nittler et al., 1997; Takigawa et al., 2018). Changing views of the dynamics and structure of the earliest solar system have provided changing contexts for such calculations in attempts to understand both early solar system solids (Cameron & Pine, 1973; Cassen, 2001; Desch et al., 2018) and condensation around giant planets and stars (Lodders & Fegley, 2002).

Results of classical calculations (figure 1) for a gas of overall solar composition (Lodders, 2020) at a given constant total pressure ($P^{tot}$) show that the most refractory elements (Al, Ca, Ti) condense as oxides at the highest temperatures (T), followed by the more abundant elements as the olivine forsterite ($Mg_2SiO_4$) and a metallic iron-rich alloy (Fe). The refractory Ca-, Al-, and Ti-rich minerals, with spinel, feldspar and melilite, form Ca-, Al-rich inclusions (CAIs). Olivine, pyroxenes, and silicate liquid form ferromagnesian chondrules in chondritic meteorites. At lower temperatures, more volatile elements condense into the existing silicates and metal, as sulfides and other minerals, and as ices at very low temperatures (Lewis, 1972a). Metal oxidation to magnetite and hydrous silicates is predicted; however, at low temperatures reactions kinetics become very important (see section "Stellar Atmospheres"; cf. Grossman et al., 2012). Indeed, the persistence of CAIs in many carbonaceous chondrites indicates a complex history of these rocks prior to their accretion.

In modern approaches, "condensation" encompasses the full range of interactions of condensates with vapor at various temperatures, pressures, or photochemical conditions, including evaporation and sublimation. These interactions can be kinetically controlled or can proceed at chemical equilibrium in closed or open chemical systems (Friend et al., 2016). Condensation calculations have been brought to bear on a variety of problems, from the origin of presolar grains (e.g., Bernatowicz et al., 1996), to exoplanet evolution (e.g., Fegley et al., 2016), to stellar atmospheres (e.g., Lodders & Fegley, 2002). Condensation provides a context for linking meteoritic and experimental observations to interpretations of spectra of astronomical objects (Takigawa et al., 2015, 2017, 2018). Recent reviews of condensation include Fegley and Schaefer (2010) and Davis and Richter (2014). Here, the primary focus is the "classical" nebula condensation problem: its history, methods applied, data considerations, experimental tests, and applications.

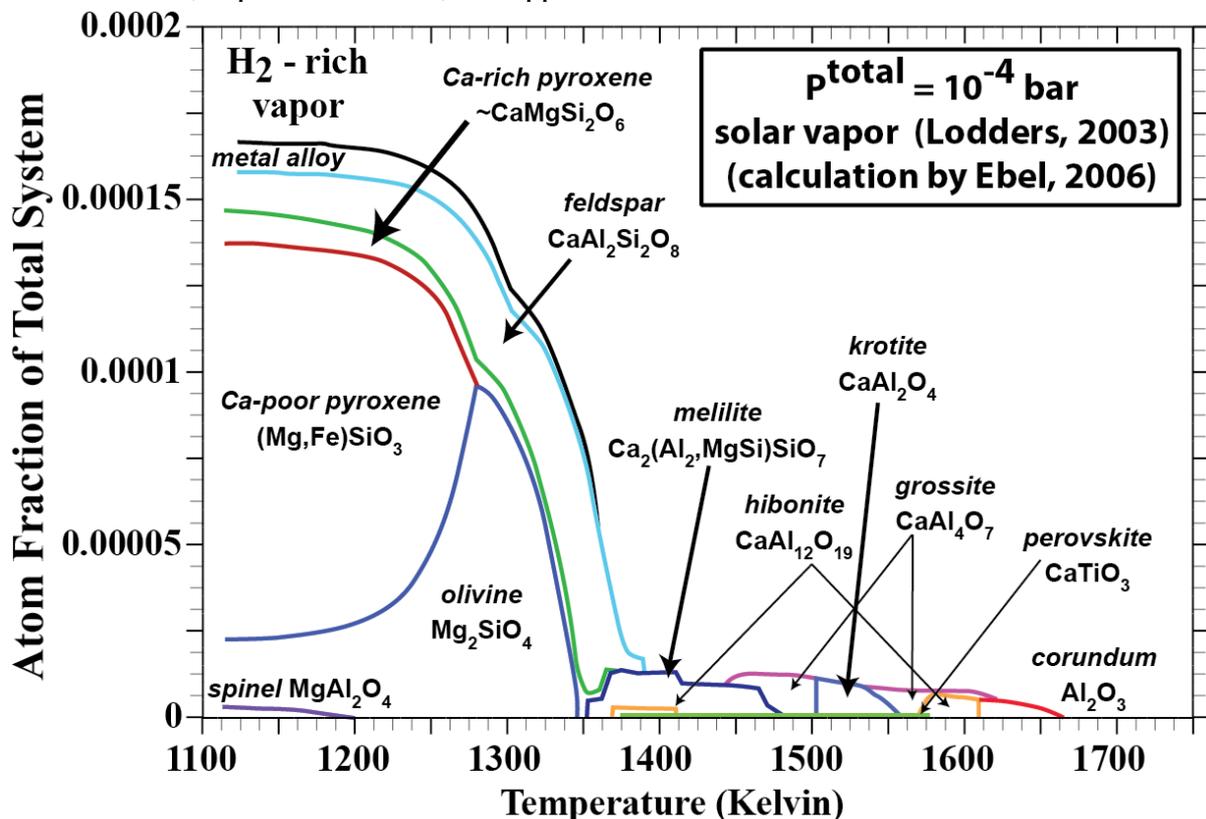

**Figure 1**: Condensation prediction (Ebel, 2006) for a vapor of solar composition (Lodders, 2003) at $10^{-4}$ bar total pressure. Cordierite $Mg_2Al_4Si_5O_{18}$ would be stable but was not included in this calculation (Ebel & Grossman, 2000).

*50% condensation temperature*

Condensation calculations for a system of solar composition yield the initial condensation temperature ($T_C$) and the T at which 50% of each element would condense ($T_{c50\%}$; Lodders, 2003; Wai & Wasson, 1977, 1979). These temperatures are the basis for understanding the depletions of volatile elements (e.g., Zn, Cl) in the Earth, meteorites, and other bodies (figure 2; Braukmüller et al., 2018; Ebel et al., 2018). However, many chemical systems differ from the canonical solar composition. For example, the enstatite chondrites are highly reduced, containing Ca- and Mg-sulfides and negligible FeO in silicates (Weisberg & Kimura, 2012). The stability of planet-forming solid materials in different regions of the Solar System, and in other solar systems, is likely to have involved significant local variations in oxidation states (Ebel & Alexander, 2011; Ebel & Stewart, 2018; Rubie et al., 2015).

**Methods**

To first order, condensation from vapor to solid and/or liquid depends on the vapor pressures of the elements that form the solid. The partial pressure of element $i$ is simply $P_i = P^{tot} * X_i$, where $X_i$ is the molar fraction of $i$ in the vapor at total pressure $P^{tot}$. Changes in $P^{tot}$ by a few orders of magnitude do not change the result in figure 1 dramatically. At higher $P^{tot}$ the condensation temperatures increase because the $P_i$ increase; with lower $P^{tot}$ they decrease. Changes in metallicity (abundance of H and He relative to everything else) have essentially the same effect as changing $P^{tot}$. Increasing metallicity increases $T_C$ by increasing the molar fraction of all the condensable elements (Ebel, 2000). Notable changes with pressure are the change of the sequence at which metal and Mg-rich olivine (forsterite) appear. At high $P^{tot}$ (such as shown in figure 1), metal starts to condense at T slightly above olivine. Around $10^{-4}$ bar, their condensation temperatures are about the same, but at lower $P^{tot}$, olivine starts to form at higher T than metal (Ebel, 2006, plate 7).

*Initial Compositions*

Condensation has been considered for a huge variety of systems, most commonly systems of "solar" or "cosmic" elemental abundance (reviewed by Lodders, 2020). This is reasonable, given that chondritic meteorites contain most rock-forming elements in near-solar relative abundances (figure 2; cf. Lodders, 2020). However, either dust enrichment or enhanced $P^{tot}$ is necessary if silicate liquids are to be stable equilibrium phases. Quenched silicate liquid is abundant as glass in chondrules in unequilibrated chondrites (Ebel et al., 2018). At the low (<$10^{-3}$ bar) pressures expected even in the dense regions of the nebula, the mole fractions of condensable elements (e.g., Si, Al, Mg, Fe) must be increased so their partial pressures increase, and they therefore condense at temperatures at which some or all of the condensed phases will melt. Dust enrichment was notably considered by Wood & Hashimoto (1993), Yoneda and Grossman (1995), Ebel and Grossman (2000; cf. Ebel, 2006, plate 10) and Fedkin and Grossman (2016). Dust enrichment (as in a nebular midplane) stabilizes silicate liquids similar in composition to those ubiquitously recorded in chondrites.

*Kinetic Models*

Some state-of-the-art simulations of protoplanetary disks currently incorporate kinetic reaction networks to determine the ionization state of the gas, including simple dust treatments (Bai, 2011; Xu & Bai, 2016), a key dynamical parameter (Bai & Stone, 2013; Gressel et al., 2015; Lesur et al., 2014; Simon et al., 2016). Reaction networks are an excellent tool for studying the evolution of gas-phase chemistry and ices with a low number of chemical species, especially in colder, more rarefied environments where non-equilibrium chemistry is expected (e.g., Aikawa et al., 2003; Gail, 1998; Herbst, 2001; Nuth et al., 2006; Semenov et al., 2004). Reaction network codes have been fully integrated with some dynamical models (e.g., Glover & Mac Low, 2007; Smith et al., 2016). However, such an approach is inappropriate for studies of the evolution of solids for comparison to detailed study of rocky Solar System samples. For example, melilite, just one mineral of interest for CAIs, has the simplified chemical formula $(Ca,Na)_2(Al_2,MgSi)SiO_7$. No feasible reaction network could track all the minerals and mineral precursors of interest in making the CAIs and Mg-silicate chondrules that are ubiquitous in chondritic meteorites.

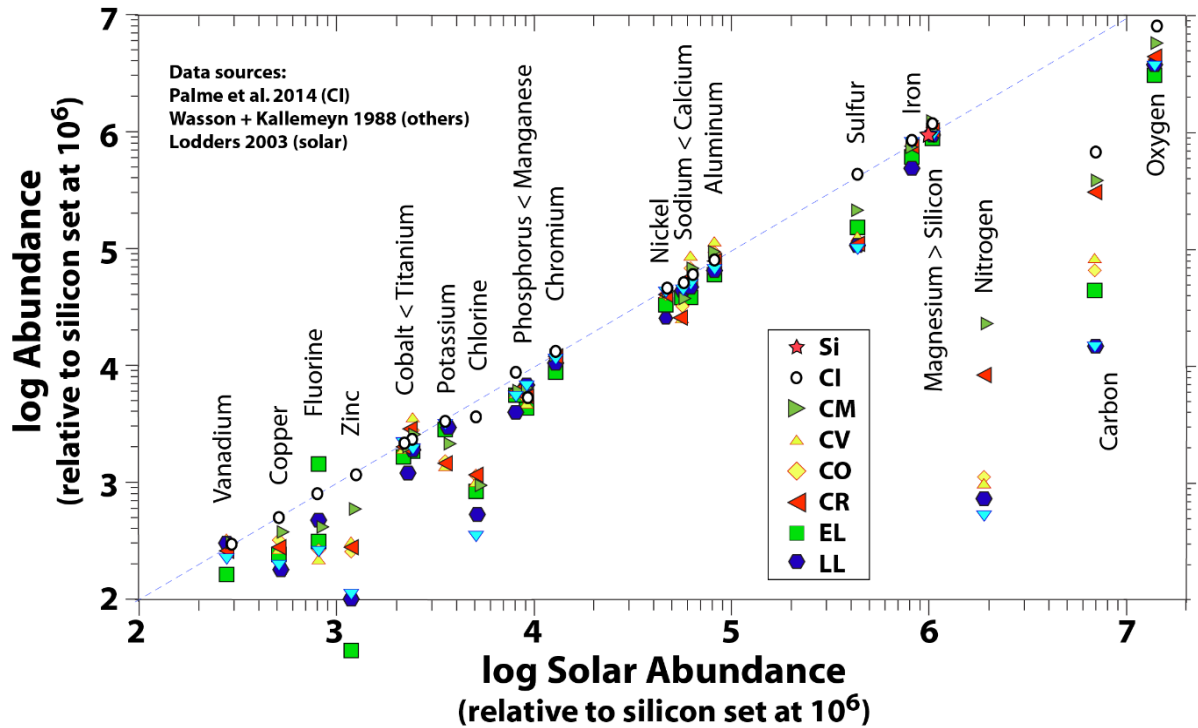

**Figure 2**: Elements more abundant than germanium in carbonaceous CI (Palme et al., 2014), CM, CV, CO, CR, enstatite (EL), and ordinary chondrites (LL), all from Wasson & Kallemeyn (1988), compared to the solar photosphere (Lodders, 2003). The CI values are canonical "solar" (Lodders, 2020). Note significant depletion of most chondrites in volatile elements Zn and Cl.

*Equilibrium Models*

      Chemical equilibrium condensation codes greatly simplify the problem at the cost of assuming local thermodynamic equilibrium at T > 1000 K. That assumption, while imperfect, is reasonable in the high-density/high-temperature regions associated with thermally processed silicates, the very regions of interest for meteoritics (Ciesla & Charnley, 2006; Wood, 1963). Indeed, similar assumptions are the basis for analysis of volcanic vent gases sampled below their equilibration temperatures (Gerlach, 1993). The assumption is also relevant to stellar atmospheres based on observations of dust formation locations (Lodders & Fegley, 1999).

      Thermochemical equilibrium calculations set the baseline for extending calculations to address kinetic effects on chemistry. Without knowing the equilibrium conditions for comparison, it is impossible to determine whether a system is out of equilibrium or to determine the magnitude of departure from equilibrium, calculated as the thermodynamic affinities of condensates, which drive chemical reactions. It has been argued, based on silica "smoke" experiments, that chemical equilibrium calculations are only relevant to secondary processes, such as annealing of clumps condensed under highly disequilibrium conditions (Rietmeijer et al., 2013); however, other experimental evidence illustrates that even short timescales of condensation produce the phases predicted by equilibrium calculations (e.g., Toppani et al., 2006).

*Stellar Atmospheres*

      The success of thermochemical equilibrium calculations in reproducing the relative abundances of gaseous species observed in stellar atmospheres and the compositions of presolar grains from stars that contributed material to our Solar System shows that chemical

equilibrium applies for stellar photospheres. The regions where equilibrium is attained have chemical reaction time scales ($t_{chem}$) that are shorter than outflow time scales ($t_{flow}$), i.e., $t_{chem} < t_{flow}$. As gas moves outward from the photosphere, $t_{flow}$ becomes small relative to $t_{chem}$, which becomes much larger with decreasing temperature. Chemical equilibrium is not reached in the low-pressure and low-temperature outer circumstellar shells where $t_{chem} > t_{flow}$. Thermochemical equilibria are frozen in at different intermediate distances between these two regions depending on where $t_{chem} = t_{flow}$ occurs for a particular reaction. Photochemistry and ion-molecule reactions driven by interstellar ultraviolet radiation determine chemical composition in the regions where thermochemistry is kinetically inhibited (e.g., Glassgold, 1996; Lodders, 2008).

If gases equilibrate in regions near the photosphere of a star, gas-solid equilibria also likely occur. However, some theoretical work implies that ideal condensation does not occur but that the gas must be supersaturated for condensates to nucleate. Prior to the discovery of presolar grains, these calculations "proved" that micron-size stardust could not form around stars. Yet, micron-size stardust exists in meteorites. Calculations assuming condensation at saturation satisfactorily reproduce (1) observed locations of dust in circumstellar shells, (2) abundance patterns recorded in presolar grains, and (3) trace element abundance patterns observed in grains from cool stars that are consistent with fractional equilibrium condensation (Amari et al., 1995, 2001; Amari & Lodders, 2007;Bernatowicz et al., 1996, 2005; Hoppe et al., 2015; Jose et al., 2004; Lodders, 2008; Lodders & Amari, 2005; Lodders & Fegley 1995, 1997a,b, 1998, 1999).

*General Considerations*

The applicability of equilibrium calculations can be tested by observing natural systems and by performing experiments. Terrestrial volcanic gases closely approach equilibrium speciation above 773 K (Symonds et al., 1994). Attempts at condensation experiments that simulate grain formation in the solar nebula or circumstellar envelopes date back over 40 years. Toppani et al. (2006) condensed crystalline olivine, noting that "None of these (previous) attempts was able to simulate the astrophysical conditions for condensation of a solar gas because of the difficulty in (1) producing a multi-element refractory gas of controlled composition and (2) condensing it at equilibrium and at high-temperature." The same criticism is valid for work dominated by $O_2$ gas, which is explosively unstable in a hot, $H_2$-rich gas (e.g., Kimura et al., 2008), or work that omits $H_2$, the major gas in a system that is solar or near-solar in composition (e.g., Kaito et al., 2003; Kimura & Kaito, 2003). Regarding the timescales needed to reach an equilibrium state, Toppani et al. (2006) noted that "under astrophysical conditions similar to those of our experiments chemical equilibrium should be attained on the timescales of about 1 hour or less." In the limit of infinite time, kinetic (reaction network) models for gas-grain chemistry should yield results identical to those of thermochemical equilibrium calculations (Young, 2007). Indeed, to properly model chemistry in a protoplanetary disk, transitions between such models would have to be seamless.

The most unequilibrated chondrites contain components (CAIs, chondrules, etc.) that are not in equilibrium with each other. The high-T minerals in CAIs would react with vapor at lower T (Figure 1), yet CAIs are present in most carbonaceous chondrites. Fractional condensation of the rare earth elements (REE) is evident even among the CAI population, a large fraction of which displays enrichment in less-refractory REE, with a complementary ultra-refractory fraction in chondrites that have overall (bulk rock) "flat" (CI-chondritic, as in the Ivuna-type chondrite, Orgueil, Lodders, 2020) REE patterns (Boynton, 1978; Ireland & Fegley, 2000). Fractional condensation calculations are a way to explore this kind of disequilibrium by removal of high-T condensates from the chemical system (Petaev & Wood, 1998, 2005). Fractional condensation also must be applied to dwarf star atmospheres where primary condensates settle within the atmosphere to form clouds (Lodders, 2004a; Visscher et al., 2010). Noting the large $MgO/SiO_2$ and FeO variations in chondrules with less variable refractory element abundances, Nagahara et al. (2005) developed a kinetic model for condensation of chondrule melts and used it to constrain cooling rates and dust enrichment. Alexander (2004) developed models for kinetic evaporation-condensation and applied them to O isotopic measurements of chondrules to constrain gas pressures and solid/gas ratios during chondrule formation.

*Nucleation and condensation constraints*

Cameron and Fegley (1982) applied classical nucleation theory (Salpeter, 1974) to estimate that several hundred degrees of undercooling are required to condense solids in the absence of nuclei, as argued by Blander and Katz (1967). However, it is likely that highly refractory interstellar or locally condensed grains that could serve as nuclei for condensation were present in many nebular regions, as calculated by Grossman et al. (2012) using the equations of Blander and Katz (1967), in a critique of Blander et al. (2004; see "Historical Review" section).

The kinetics of condensation have not been subject to the same experimental scrutiny as kinetics of evaporation, in part due to the difficulty of experiments. Tachibana and Takigawa (2013) compared the kinetics of metallic iron and forsterite ($Mg_2SiO_4$) condensation, finding very strong differences in their condensation coefficients, indicating that formation of metallic dust is much more efficient than formation of forsterite dust in circumstellar environments (cf. Tachibana et al., 2011).

*Other Energetic Effects: Spinel*

Many have noted the ubiquitous presence of $MgAl_2O_4$ spinel in the cores of CAIs in carbonaceous chondrites, often intergrown with hibonite (figure 1; MacPherson, 2014). Hibonite laths enclosed by aggregates of spinel with perovskite inclusions, surrounded by Ca-rich pyroxene, were interpreted by Han et al. (2015) as disequilibrium condensates. Their careful nanoscale study showed epitaxial nucleation and growth of spinel on hibonite, which they interpreted as evidence of a lower activation energy for spinel formation compared to more thermochemically stable melilite (figure 1). Abundant stacking defects in hibonite, a mineral that includes spinel-like blocks (Nagashima et al., 2010), would make replacement of hibonite by spinel energetically favorable. A complete equilibrium calculation would have to account for these energies, thermodynamics being an exercise in accounting (R. O. Sack, personal communication, c. 1987). Modern analytical methods (e.g., Bolser et al., 2016; Han & Keller, 2019) reveal microstructural evidence of energetic phenomena that are not accounted for in existing thermochemical condensation calculations that are based solely on macroscopic thermochemical equations of state.

## Condensation Calculations

Calculation of the equilibrium state of a chemical system requires the calculation of the global minimum in chemical potential energy, analogous to balls rolling downhill to minimize gravitational potential energy. In that analogy, impediments to ball motion would constitute kinetic barriers to reaching equilibrium. In principle, the energy of a chemical system could be minimized (i.e., at chemical equilibrium) as a function of any two extensive variables (such as entropy and volume, variables that depend on the amount of matter present) or intensive variables (such as temperature and pressure, variables that are independent of the amount of matter in the system) holding the other two variables constant, using the appropriate choice of state function: internal energy, enthalpy, and Helmholtz or Gibbs free energy (Ghiorso & Gualda, 2015; Ghiorso & Kelemen, 1987). In practice, most equilibrium techniques minimize the Gibbs free energy ($G_{system}$), subject to mass balance constraints. This makes sense given that most of the thermodynamic data for solids and liquids, and some gaseous species, have been obtained over the past century by experimental petrologists and materials scientists at controlled T and P conditions. Mass balance constraints require that the calculated equilibrium abundances of chemical species always return the elemental abundances input to the computation.

White et al. (1958) are widely acknowledged (e.g., Eriksson, 1971) to have first described a method of free energy minimization for a gaseous system and its implementation in computer code. They calculated rocket fuel combustion in C-H-O-N systems using the method of Lagrange multipliers, or 'stoichiometric' method, using a simplex (first-order) technique. They did not include condensed species, but they remarked that inclusion of such species would be a simple and straightforward extension of their technique. Boynton (1960) presented a method for polyphase systems, including nonideal condensed solutions with known activity coefficients.

The difficulty of computing equilibrium, even in an ideal mixture of gases that obey the ideal gas law, is that as the number of elements increases, so do the numbers of mixed species. An H-O vapor might contain H, O, $O_2$, $H_2$, $H_2O$, and OH species. A more complex system Al-O-H could include gaseous species Al and AlO. Each species, in terms of the basis elements, is formed by a reaction, such as $2H + O = H_2O$, $2O = O_2$, or $Al + O = AlO$. The Gibbs free energies ($G_{rxn}$) of these reactions may be calculated from tables (e.g., Chase, 1998). Constrained by mass balance equations, such as $H^{total} = 2H_2O + 2H_2 + OH + H$, the free energy of the gaseous system, $G_{gas}$, can then be minimized as a function of the $G_{rxn}$ of all the independent species formation reactions. In the Al-O-H example, a system of three elements has three mass balance equations and five speciation reactions. Eight equations can then be solved for the amounts of eight gaseous species present at a fixed T and P.

Solid $Al_2O_3$ (corundum) is stable relative to the gas at a particular P and T if $G_{rxn}$ for $2Al_{(g)} + 3O_{2(g)} = Al_2O_{3(s)}$ is negative. That is, if condensation of $Al_2O_3$ reduces the free energy of the system. Computing how much of the solid will condense requires computation of the first derivative of $G_{gas}$ with respect to changes in Al or O in the gas. This is a computationally difficult task in a system with more than a few elements (Ebel et al., 2000). A closed form, polynomial expression for $G_{gas}$ would facilitate such a calculation, given that the equations of state for pure solids and solid solutions are already represented by polynomial expressions (e.g., Sack & Ghiorso, 1989, 1994).

Recent efforts use tools for manipulating algebraic expressions (e.g., Python SymPy) to build a closed form equation expressing $G_{gas}$ as a function of T, P and the mole fractions $X_i^{tot}$ of the basis species $i$ (Boyer et al., 2019, 2020; ENKI, 2020). Usually the basis species are gaseous monatomic elements - Al, H, O in the example - where the sum of $X_i^{tot}$ in the gas is 1. The molar Gibbs free energy of the gas $\bar{G}_{gas}$ is then the sum of the molar $G$ of all the gaseous species $k$, each expressed in terms of the basis species. For the simple example of H and O, with species $O_2$, $H_2$, $H_2O$, and OH, we have $X_H^{tot} + X_O^{tot} = 1$. Defining the bulk composition of the gas by $r_1 = X_O^{tot}$, so $X_H^{tot} = 1 - r_1$, the species are simply "ordered" combinations of the endmembers O and H, along the lines of an associated solution model (Kress 2000, 2003). Let $s_1 = Y_{H2O}$, $s_2 = Y_{OH}$, $s_3 = Y_{O2}$, $s_4 = Y_{H2}$ represent the abundances of these ordered species. Then the abundances of *all* the species (including H and O) must be normalized by their sum, which is:
$$Y_H + Y_O + Y_{H2O} + Y_{OH} + Y_{O2} + Y_{H2}$$
$$= [r_1 - s_1 - s_2 - 2s_3] + [1 - r_1 - 2s_1 - s_2 - 2s_4] + s_1 + s_2 + s_3 + s_4 = 1 - 2s_1 - s_2 - s_3 - s_4 = M$$
The mole fractions of the species are then
$$X_O^{sp} = [r_1 - s_1 - s_2 - 2s_3]/M,\ X_H^{sp} = [1 - r_1 - 2s_1 - s_2 - 2s_4]/M,\ X_{H2O}^{sp} = s_1/M,\ \text{etc.},$$

and
$$\bar{G}_{gas} = \sum_k^6 X_k^{sp} \bar{G}_k^o + RT[X_k^{sp} ln X_k^{sp}],$$
where $\bar{G}_k^o$ are standard state molar Gibbs energies of formation of species $k$, and $R$ is the gas constant.

At chemical equilibrium among all the gas species, the change in $\bar{G}_{gas}$ with abundance of each of the four species $i$ is zero:
$$\partial \bar{G}_{gas} / \partial s_i = \sum_k^6 \left\{ \left( \partial X_k^{sp} / \partial s_i \right) \bar{G}_k^o + RT \left[ \partial (X_k^{sp} ln X_k^{sp}) / \partial s_i \right] \right\} = 0$$
where application of the chain rule yields four equations
$$0 = \sum_k^6 \left( \partial X_k^{sp} / \partial s_i \right) \{ \bar{G}_k^o + RT(ln X_i + 1) \}$$
and the relevant derivatives can be obtained for a system of arbitrary complexity using a symbolic algebra code (e.g., SymPy, Wolfram Alpha, MATLAB). A solver can then be used to solve for $s_1$ - $s_4$ given an input value for the independent compositional variable $r_1$. This is a fundamentally different approach from those of White et al. (1958) and Ebel et al. (2000). It is extensible to a multi-element gas with hundreds of possible gaseous species (Boyer, 2020).

Larimer (1967), Grossman (1972), Lodders (2003), Fegley and Schaefer (2010) and many others have ably presented the basic outlines of the classical approach to condensation calculations in $H_2$-rich nebular environments. Reviews of numerical methods were published by Zeleznik and Gordon (1968) and van Zeggeren and Storey (1970), the latter describing both first- and second-order polynomial fits to the Gibbs free energy surface. Both also reviewed particular applications, primarily to gas-phase equilibria related to metallurgical and chemical engineering problems (e.g. Oliver et al., 1962). Gautam and Seider (1979) and numerous other workers have treated gas-liquid, gas-solid and other systems in the chemical engineering literature. Gill et al. (1981) addressed the problem from the standpoint of numerical optimization (i.e., as a multivariate minimization problem). Connolly and Kerrick (1987) reported algorithms for calculating stable phase equilibrium topologies in the CALPHAD framework, which will not be further explored here. Wood and Hashimoto (1993) reviewed methods including publications from the Soviet Union. Specific approaches are further addressed here in the section "Historical Review".

*Evaporation*

Evaporation, more properly sublimation, is a sort of "kinetic de-condensation, requiring the calculation of the vapor pressure of the evaporating gaseous species over a solid or liquid. Such a calculation requires exactly the same kind of algorithm as for condensation calculations. Series of experiments in Japan (e.g., Hashimoto, 1983; Tachibana et al., 2002; Tsuchiyama et al.,1999) and Chicago (e.g., Richter et al., 2002, 2011; Wang et al., 2001) have yielded evaporation coefficients that allow application of the Hertz-Knudsen equation for evaporative flux from molten droplets as a function of surface area, time, and vapor pressure (Richter et al., 2002). Application requires knowledge of the chemical activity of evaporating species in the melt, as is also required for calculating condensation of melt. Derived evaporation coefficients are highly sensitive to the activity model adopted for the melt (Ebel, 2005).

**Thermochemical Data**

Thermochemical computations require an accurate and self-consistent set of thermodynamic data describing equations of state for all gaseous, solid, and liquid compounds present, which determine their internal energy (Anderson & Crerar, 1993).

Popular data sources are the third and fourth editions of the National Institute of Standards & Technology - Joint Army Navy Air Force (NIST-JANAF) thermochemical tables (Chase, 1998; Chase et al., 1985) or the Russian counterpart made by the Institute of High Temperatures of the Russian Academy of Science (Glushko et al., 1991; Gurvich et al., 1991). Many calculations (e.g., Burrows & Sharp, 1999; Yoneda & Grossman, 1995) directly utilize regressions of the Gibbs free energy of formation from the elements as tabulated in compilations (e.g., the JANAF tables, which are also the basis of the "NASA-polynomials"; McBride et al., 1993). However, these tables have been shown to contain errors in the quantities derived from heat capacity and 298 K standard enthalpies and entropies (Ebel & Grossman, 2000; Lodders, 1999, 2004, see also appendix in Lodders 2002), and undiscovered errors and inconsistencies may remain.

A more convenient approach for condensation (e.g., Ebel et al., 2000) is to apply the so-called "apparent" $G$ of formation of compounds from the elements as the difference between the compound's absolute $G$ at T and P and the constituent elements at $T_{ref}$ = 298.15 K and $P_{ref}$ = 1 bar (Anderson & Crerar, 1993, their Section 7.4.2). This approach avoids discontinuities in the high-T standard state properties of the elements, a source of errors in calculation of derived quantities in the JANAF tables (Lodders, 1999). It is simplest to convert traditional $G$ of formation at the T of interest to apparent $G$ of formation using polynomial representations of the Giauque functions for the elements at their standard states at 298.15 K and $P^{tot}$ = 1 bar ($10^5$ Pa) (Anderson & Crerar, 1993, their Eq. 7.7; Berman, 1988). The traditional $G$ of formation at the T of interest is first computed from a compound's standard state enthalpy and entropy of formation from the elements at 298 K and $P^{tot}$ = 1 bar and a regression of its heat capacity at 1 bar ($C_P$) in terms of temperature (T), accounting for phase transitions. In this way, all compounds are referenced to the same set of elements in their reference states for internal consistency. A similar approach was taken by Fegley and Lodders (1994), Fegley et al. (2016), Costa et al. (2017), and Sossi and Fegley (2018).

Regression coefficients for $C_P$ are provided in e.g., Knacke et al. (1991), Robie et al. (1979), and Berman (1983, 1988), or may be calculated with some labor from tables of $C_P$ - T data (e.g., Chase, 1998). Other sources include: Pedley and Marshall (1983), Mills (1974), Ackermann and Chandrasekharaiah (1975), and sources listed in Lodders and Fegley (1993). The Cp-functions in Knacke et al. (1991) and older compilations (e.g., Hultgren et al., 1973) do not always return the correct heat capacity data, so the parameters provided must be checked against the tables and the original literature.

In the global literature on mineral thermodynamic data, there are several primary sources and a great variety of secondary sources. For more obscure phases, such as hibonite, it is unlikely that 'absolutely true' endmember properties can be established. Several phases are important condensates at high temperature, among them corundum ($Al_2O_3$), hibonite (Ca-hexaluminate), grossite (Ca-dialuminate), krotite (calcium aluminate), pure spinel, åkermanite and gehlenite. The role of choices of database and a comparison of calculations for these minerals is presented by Ebel (2006, his Figure 1).

Particular minerals have great importance in condensation calculations, for example there are several determinations of the stability of sodalite $Na_4(AlSiO_4)_3Cl$, a mineral critical to condensation of chlorine (compare Stormer & Carmichael, 1971, to Komada et al., 1995). Fegley and Schaefer (2010) reported a $T_{c50\%}(Cl)$ of 400 K, a revision from the 948 K calculated by Lodders (2003), who used older data for sodalite. Wood et al. (2019) obtained $T_{c50\%}(Cl)$ of 472 K using a much older code (Wood & Hashimoto, 1993). Persistence of chlorine in the vapor phase lowers the $T_{c50\%}$ for volatile lithophile and siderophile elements.

In Table 1, the reference values for molar heat of formation at 298 K ($\Delta H_{fm, el}$) from the elements and the standard entropy (S°) and heat capacity ($C_P$) at 298 K for hibonite and grossite are compared across several references. The differences between data compilations are larger than the uncertainties cited for the various phases. Differences between primary experimental data sources are problematic for phases like hibonite, for which few such data exist.

Recent efforts to "rationalize" the vast experimental and theoretical literature on equations of state include "Active Thermochemical Tables" (Ruscic & Bross, 2019; Ruscic et al., 2004). This work is potentially promising but limited to 14 elements. It is difficult to imagine this approach, as well as the CALPHAD approach, as being capable of addressing the complex, non-linear behavior of complex solid solutions like clinopyroxenes (Sack & Ghiorso, 1994) or melilites (Beckett & Stolper, 1994; Grossman et al., 2002). The IVTANTHERMO project (Belov et al., 1999, 2018), while comprehensive, does not touch on the thermodynamics of solid solutions, nor on questions of vapor-liquid equilibria (Belov, 2015).

**Table 1:** Comparison of standard state properties of hibonite ($CaAl_{12}O_{19}$) and grossite ($CaAl_4O_7$).

| Hibonite | dHf-298 K (kJ) | err (kJ) | S0-298 | method |
|---|---|---|---|---|
| Glushko 1979* | -10742.8 | 12.9 | | reference is to a compilation |
| Eliezer et al. 1982* | -10605.2 | 10 | 424* | theory: phase diagram analysis |
| Hemingway 1982* | -10813 | 20 | est. 432 | theory: corresponding states |
| Berman 1983 | -10735.3 | n/a | 360.98 | theory: internally consistent CMAS database |
| Geiger et al. 1988 | -10722 | 12 | | Alkali-borate solution calorimetry at 1063 K. |
| **Grossite** | | | | |
| Skolis et al. 1981* | -4011.2 | 7.5 | | From high-T electrochemistry, extrapolated lower. |
| Glushko 1979* | -4029.6 | 5.3 | | reference is to a compilation |
| Eliezer et al. 1982* | -3995.3 | 3 | | theory: phase diagram analysis |
| Hemingway 1982* | -4023.8 | 4.6 | | From HF calorimetry of Koehler et al 1961. |
| Berman 1983 | -4006.7 | | 175.06 | theory: internally consistent database |
| Geiger et al. 1988 | -4007 | | | Alkali-borate solution calorimetry at 1063 K. |

Columns give the molar heat of formation at 298 K ($\Delta H_{fm, el}$) from the elements, cited uncertainty in $\Delta H_{fm, el}$, and the standard entropy (S°) at 298 K. Some values (with *) are quoted directly from Geiger et al. (1988).

While calculations treating melt-solid equilibria applicable to planetary interiors (e.g., Ghiorso & Sack, 1995; Wood & Holloway, 1984) must explicitly account for the effect of $P^{tot}$ on phase stability, the effect of P on condensed phase stability may be neglected at $P^{tot} <$ 1000 bar. However, the effect of $P^{tot}$ on the free energies of gas species cannot be neglected.

Internal consistency among data sets is important for accurate calculations of equilibrium chemistry. As much as feasible, equations of state used in such calculations should be from single sources (e.g., Berman, 1988; Robie et al., 1979) or referenced to a single set of thermodynamic data for the elements (as in Ghiorso & Sack, 1995).

*Solid Solutions*

Condensation calculations intended to address subtle variations in the compositions of condensed solids, for example Mn-bearing olivine, $(Mg,Fe,Mn,Ca)_2SiO_4$ (Ebel et al., 2012), require careful application of solid (and liquid) solution models. Internal consistency and reliance on extensive experimental calibrations are essential. For some solid solutions, such as melilite, $Ca_2(Al_2,MgSi)SiO_7$, existing models are simply inadequate (Grossman et al., 2002, their Figure 1; Ustunisik et al., 2014). Determination of solid solution models for complex phases (e.g., Ca-rich pyroxene) consistent with observed data (e.g., lattice site occupancies, atom exchange experiments, phase equilibria, thermal and spectral properties) is a difficult task (e.g., Sack & Ghiorso, 1989, 1994). Such models must be consistent with models for coexisting solutions (e.g., olivine, spinel) and silicate liquids to effectively constrain natural phenomena such as melt crystallization (Ghiorso & Sack, 1995). Equilibria of vapor with condensed melts and solid solutions adds another level of complexity (Ebel et al., 2000).

Calculation of the condensation behavior of trace elements or exotic solid solution phases provides a particular challenge. Lodders (2003) has addressed the former in calculating the 50% condensation T ($T_{c50\%}$) for the entire periodic table using activity expressions where experimental constraints were available. Exotic phases such as Na-bearing kosmochloric Ca-rich pyroxenes (Joswiak et al., 2009) or osbornite (Ti,V)N (Brownlee, 2014) from comet Wild 2, scandium-rich minerals in CAIs (Ma et al., 2014), or $NaCrS_2$ caswellsilverite (Weisberg & Kimura, 2012) in enstatite chondrites lie outside the realms explored by most terrestrial experimental petrologists. These are challenges for the future.

## Experimental Tests of Condensation

Experimental work at the high temperatures required to explore refractory silicates is difficult. Kushiro and Mysen (1991) determined vapor-solid-liquid stability fields for forsterite, periclase, enstatite and silica in the $MgO-SiO_2-H_2$ system. Tsuchiyama and Kitamura (1995) constrained solid-gas equilibria in the Fe-Mg-Si-O-C-H system to explore redox states in chondrites. Lauretta et al. (1997) investigated the nebular sulfidation of iron. Davis and Richter (2014) have reviewed much of the previous experimental work on systems relevant to cosmochemistry, particularly evaporation.

### Direct Condensation

Kushiro and Nagahara (1988; cf. Kushiro & Mysen, 1991) demonstrated by experiment that major chondrite-forming minerals, and not amorphous grains, condense directly from vapor of solar composition in the order calculated by Grossman (1972) at low $P^{tot}$. Toppani et al. (2006) performed direct condensation of crystalline oxides and silicates, including forsterite, at $4 \times 10^{-3}$ bar from 1318 to 1558 K for runs of 4 to 60 minutes, attaining steady-state gas-solid equilibria in < 1 hour. They found that kinetic processes favor specific phases, such as $MgAl_2O_4$ spinel. Takigawa et al. (2019) recently formed alumina like that found as presolar grains, simulating condensation conditions around asymptotic giant branch (AGB) stars.

### Vapor-Free Tests

Experimental verification of condensation calculations has revealed deficiencies in models for liquid and solid solutions, but it has also confirmed their general applicability. Ustunisik et al. (2014) directly tested the identities of minerals predicted by Ebel (2006) to condense at particular $T_o$, $P_o^{tot}$ and bulk composition $X_o$ (C1-dust enrichment; figure 3a), omitting metal alloy. Because the difference in chemical free energy of Ca, Mg, etc. is large between vapor and condensed phases, they assumed that the amounts of Ca, Mg, etc., predicted to condense with oxygen at any $T_o$, $P_o^{tot}$ and $X_o$ were correct. Because the free energy surface of the condensed liquid + mineral assemblage is nearly flat, i.e., the differences in free energies among condensates are small, they expected to find errors in prediction of the identities, compositions, and amounts of minerals stable in the liquid. Indeed, they found that melilite, not hitherto predicted to be stable in liquid-bearing condensate assemblages (figure 3a), is in fact a stable phase in certain P, T, X regions (figure 3b). Since melilite is common in igneous (melted) Type B CAIs (MacPherson, 2014), this is comforting, but the result indicates that we do not properly understand the thermodynamic behavior of the melilite solid solution, as demonstrated by Grossman et al. (2002, their Figure 1).

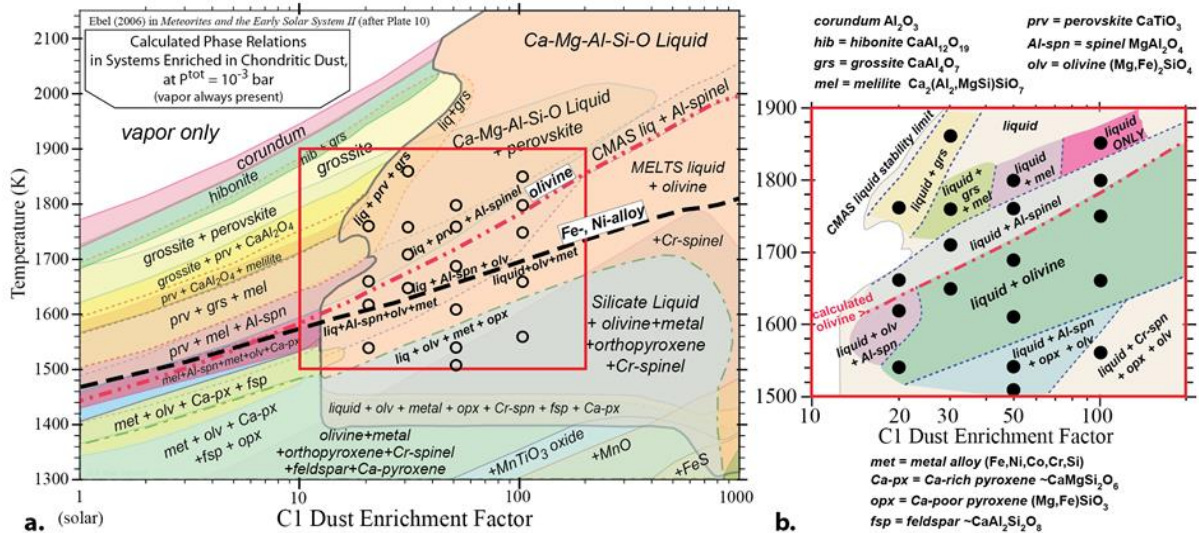

**Figure 3**: Condensation predictions (a) and experimental results (b). Red dash-dot line indicates the T below which olivine is stable, black dashed line the T below which metal alloy is stable. Black dots in (b) indicate experiments at composition and T of condensed oxide assemblages of (a), omitting metal alloy. (a) Adapted from Ebel (2006, Plate 10), (b) Adapted from Ustunisik et al. (2014).

## Historical Review

Grossman (1972) reviewed the history of prior work on the nebular condensation problem. Work before the advent of the digital computer required a variety of simplifying assumptions. Urey (1951) began considering chemistry, in the form of fundamental chemical reactions, in the context of Solar System formation that had until then been primarily a problem of physical modeling (e.g., Kuiper, 1951). Latimer (1950) addressed the formation of the Earth from an isolated cloud of 'solar' composition by considering the thermodynamic stability of simple silicates and metal in an $H_2$-dominated vapor. Kamijo (1963) calculated saturation vapor pressures in M-type circumstellar envelopes, concluding that they could contribute solid silica ($SiO_2$) to the interstellar medium. In investigations of opacity sources in nebulae and stars, Gaustad (1963) considered a few major condensate species. Tsuji (1964, 1973) calculated the condensation of graphite from H-C-N-O stellar atmospheres.

Wood (1963) calculated the FeO content of silicate liquid in equilibrium with a solar gas and calculated vapor-solid equilibria in the H-O-Si-Mg-S-Fe system using the then-new JANAF thermochemical tables (Chase, 1998). His P-T diagram presented an "ill defined" (sic) liquid-solid boundary at $P^{tot} > 100$ atm. Wood's investigation of condensation became the foundation for later studies in the petrological context of chondritic meteorites.

Mueller (1964) used thermochemical data to calculate the 'atmosphere of crystallization' of chondritic meteorites using the Fe-Mg-$O_2$ balance between olivine, pyroxene and gas. At constant activity of Fe in metal, $X_{FeO}$ in silicates increases with $f(O_2)$ in the gas, and $X_{Ni}$ in metal increases, at constant silicate composition. Equilibria of H and C were also considered, with CO dominating (not $CO_2$), and the ratio of partial pressures $H_2O/H_2 < 1$, in systems in equilibrium with bodies of chondritic composition. The results were used to constrain possible environments of chondrite crystallization, how their textures were produced, and the stability of hydrocarbons. A 'full equilibrium' calculation was not performed. Mueller (1965) calculated the stability of sulfur compounds on Venus.

Lord (1965) summarized earlier thermodynamic calculations of vapor speciation in stellar atmospheres. He calculated equilibria in a solar gas between 1700 and 2000 K for 1 and $5 \times 10^{-4}$ atm., considering 150 species of pure condensate, including sulfides. Tables presented the results at 1700 and 2000 K and showed the $p(H_2)$ required for condensation of each species. Solid species $Al_2O_3$, W, $ZrO_2$, and $MgAl_2O_4$ condensed at 2000 K.

Blander and Katz (1967) discussed condensation kinetics, predicting the relatively rapid nucleation of liquids relative to solids, and silicate liquids relative to metallic liquids, assuming the absence of preexisting nucleation sites. Since surface energies of liquids are much smaller, barriers to nucleation are lower than for solids, and metastable formation of silicate liquid below its solidus would be likely. The high surface tension of iron liquid relative to silicate liquid means that a higher degree of supersaturation would be necessary for nucleation of iron liquid. However, Grossman et al. (2012) later disproved the basic assumption of this work.

Larimer (1967) considered fast and slow cooling of a solar gas at $P^{tot} = 6.6 \times 10^{-3}$ atm, conditions corresponding to the T-independent $P^{tot}$ at the asteroid belt (2.8AU) in Cameron's (1962) two solar mass, radiative equilibrium nebula model. If diffusional equilibrium is maintained during slow cooling, each element condenses over a wider temperature range. Larimer's method, very well described, was inexact in some ways, but quite adequate for his purpose of determining relative condensation temperatures of the elements. Larimer included many elements, but only the few most important gas and solid species for each element. The effects of solid solution were considered only in terms of mechanical mixing in alloy, sulfide, oxide, and silicate species. Disequilibrium effects, particularly diffusion of trace metals into metal alloy grains, were discussed. Larimer and Anders (1967) used the calculations of Larimer (1967) to conclude that most chemical fractionations observed in meteorites can be explained in terms of the accretion temperatures and volatilities of condensed elements, the so-called "two component model" (Alexander, 2019; Anders, 1964; cf., Wasson,1977 for a contrary view).

Shimazu (1967) calculated condensation for the system H-O-C-N-S-Si-Mg-Fe at varying $f(O_2)$ in exploring formation of the Earth. Circumstellar condensation was addressed by Donn et al. (1968) and Fix (1970). Gilman (1969) included liquid condensates in calculating circumstellar grain condensation for 12 elements, at $P^{tot}=5$ Pa ($5 \times 10^{-5}$ bar), C/O atomic ratios 0.4 to 2.5, at T from 1200 to 2000 K. Grossman (1972) remarked that "these were full equilibrium calculations in the sense that all species, gaseous and condensate, were relaxed to their equilibrium abundances for a given temperature and total gas pressure." Gilman's results differ from others in finding $Al_2SiO_5$ stable at ~1700K at C/O ~0.4 (~solar).

Arrhenius and Alfven (1971) considered low-density, partially excited gas in which the gas temperatures are much higher than the solid grain temperatures. They concluded that "the materials in meteorites have properties which in some cases permit, in others strongly suggest or require the assumption that they are primary and largely unaltered solids grown in extreme thermal disequilibrium with the surrounding gas phase." Their arguments are largely textural, based on condensation experiments by themselves and others.

Lewis (1972a, b) investigated equilibrium (homogeneous) condensation, in which "chemical processes are rapid compared to the rate of accretion", and also fractional (inhomogeneous) condensation, in which "accretion proceeds nearly instantly". Lewis's primary goal was to understand the compositions of the outer planets and their satellites, the differing densities of the planets, and the metal-silicate fractionation expressed in L- and H-group chondrites. Lewis' work (1973) informed a long-standing paradigm of planetary composition controlled by thermal gradients with heliocentric distance, which was overthrown by chemical measurements from the MESSENGER mission (Ebel & Stewart, 2018).

Grossman (1972; cf. Grossman & Larimer, 1974), was the first to perform large-scale calculation of the high temperature condensation sequence of a gas of solar composition, at $P^{tot} = 10^{-3}$ atmospheres, with rigorous attention to mass balance at each step. His calculation was for the elements H, O, C, N, Mg, Si, Fe, S, Al, Ca, Na, Ni, P, Cr, Mn, F, K, Ti, Co and Cl between 1200 and 2000 K. He also explored the effect of $P^{tot}$ on iron and olivine condensation, concluding that at all $P^{tot} > 7.1 \times 10^{-5}$ atm, metallic iron condenses at a higher T than forsterite. His method was similar to that of Larimer (1967), written in FORTRAN for an IBM model 360/65 computer using punched cards. Operator intervention was required between temperature steps. Kushiro and Nagahara (1988) verified Grossman's (1972) theoretical results in broad outline by experiments in the Mg-Fe-Si-O-H system at 1773 and 1823K at low $P(H_2)$.

  Grossman (1973) added refractory trace elements (Os, Sc, Re, Ta, Zr, W, Y, Hf, Mo, Ru, Ir, V, Th) and considered the formation of their gaseous oxides (ScO, TaO, ZrO, $ZrO_2$, WO, YO, HfO, $HfO_2$, MoO, VO, ThO, $ThO_2$) and pure crystalline oxides ($Sc_2O_3$, $Ta_2O_5$, $ZrO_2$, $Y_2O_3$, $HfO_2$, $MoO_3$, VO, $VO_2$, $V_2O_3$, $ThO_2$). He noted that refractory nitrides and carbides of these trace elements may condense at even higher T than oxides, and that divalent Eu would differ in behavior from the primarily trivalent REE. Data for many species were lacking for these early calculations. Grossman and Clark (1973) calculated upper limits to nebular $P^{tot}$ from the stability fields of refractory CAI phases. Grossman and Olsen (1974) included a model for Fe-Ni-Co-Cr-Mn alloy in condensation. The REE occur as monatomic or monoxide vapor species, but Grossman and Ganapathy (1976) revisited the REE, including $CeO_2$ as a gaseous species, calculating $CeO/CeO_2 > 1$ for solar composition vapor, an early recognition of the importance of redox state.

  Boynton (1975) inserted trace elements (REE) into high-T major element condensates given the stability of those pure condensates relative to gas. He did not consider solid solution of major elements such as olivine, $(Mg,Fe)_2SiO_4$. The REE were not part of the minimization but were partitioned among calculated solid and gas phases according to estimates of their distribution coefficients. This essentially applied the same technique Larimer (1967) had applied to metals, to improve on the work of Grossman (1973) who had considered the condensation of refractory trace elements into pure phases. Boynton (1978) focused on REE volatility as a function of oxygen fugacity.

  Larimer (1975), updated by Larimer and Bartholomay (1979), used a code developed by Reese (1973) following Gordon and McBride (1976, 1994; McBride & Gordon, 1996; cf., Huff *et al.*, 1951) to investigate the effects of C/O ratios from ~0.4 to 1.4 on silicate $T_C$ in a vapor of otherwise solar composition to which C was added. They included H, C, N, O, Mg, Al, Si, S, Ca, Ti, and Fe (but not solid TiC). They discovered strong decreases in the $T_C$ of oxides and silicates above C/O ~ 0.9, but no effect on Fe metal condensation. Larimer and Bartholomay (1979) cited errors of ±30°C in the predicted condensation temperatures. They reported results holding O constant (solar) while C varied, stating that the opposite case is "nearly identical, with a few notable exceptions". Kelly and Larimer (1977) considered 15 siderophile (iron-loving) elements to infer accretion temperatures of iron meteorite parent bodies. Wagner and Larimer (1978, cited by Wark, 1987) investigated the stability of refractory oxide liquid in solar gas and reported marginally stable liquid as the highest T condensate from 1 to $10^{-6}$ atm.

  Palme and Wlotzka (1976) calculated the gas - solid condensation behavior of W, Os, Re, Mo, Ir, Ru, Pt, Rh, Co, Pd, and Au to explain refractory metal grains in CAIs (El Goresy et al., 1978). They noted the lack of knowledge of activity coefficients in refractory metal alloys. Lattimer et al. (1978) refined solar composition calculations and calculated adiabatic condensation from gas compositions predicted in supernova shells for the first time, in the context of recently discovered isotopic anomalies in meteoritic oxygen and magnesium. Distinct "shells" are computed and observed to exist in supernovae, where nuclear reactions produce different mixtures of elements and isotopes in each shell through nucleosynthesis.

McCabe et al. (1979) considered the condensation of SiC to explain low SiO and SiS gas abundances observed in the dying red giant star IRC+10216, by matching chemical equilibrium calculations at T=1250K, P=100 dyne/cm$^2$ (10 Pa, 10$^{-4}$ bar)

Davis and Grossman (1979) calculated condensation behavior of REE to explain REE group II and III patterns in CAIs (Boynton, 1975; MacPherson, 2014). Their two-component model for REE removal has REE in ideal solution in perovskite CaTiO$_3$, and in another solid uniformly enriched in REE. They concluded that perovskite was removed over a very narrow T range, multiple refractory element-bearing components were present, and that gas/solid REE fractionations occurred under conditions *more* reducing than a solar gas (Beckett, 1986).

Barshay and Lewis (1978) studied gas speciation along a calculated adiabat in the deep atmosphere of Jupiter, with > 500 compounds of 27 elements, using "an elaborate computer program" (Barshay & Lewis, 1976) to compute fugacities. They calculated the stability of condensates by hand, to report upper limits on the equilibrium abundances of gas species during depletion by condensation. Lewis and Prinn (1980) showed that the condensation of methane and ammonia in cool nebular regions was inhibited by the slow rate of reduction of CO and N$_2$ in the early solar system, with significant effects on the composition of cometary ices and of the Jovian planets, and on outgassing of ice-rich satellites. They built on earlier kinetic arguments regarding CO reduction in Jupiter's atmosphere (Prinn & Barshay, 1977).

Sears (1978) calculated the P-T conditions for Ni, Ga and Ge condensation into metal precursors of iron meteorites. Sears (1980) reviewed condensation and accretion models for the formation of highly reduced enstatite (E) chondrites, and computed condensation using new data for the activity coefficient of Si in Fe metal. Sears concluded: "The reduced species in E chondrites and aubrites were isolated from the nebula soon after their formation", and "if the nebular pressure was on the order of 1 atm, the composition may have been near to cosmic."

Saxena and Eriksson (1983) adapted the proprietary program SOLGASMIX with the addition of regular solution models for potential condensates, including (Mg,Fe)Al$_2$O$_4$ spinel, (Mg,Fe,Ni)Si$_{0.5}$O$_2$ olivine, (Mg,Fe,Ca$_{0.5}$Mg$_{0.5}$,Al$_2$)SiO$_3$ orthopyroxene, (Mg,Fe,Ca)Al$_{0.67}$SiO$_4$ garnet, Ca$_2$(Al$_2$,MgSi)SiO$_7$ melilite, (Fe$_3$,FeAl$_2$,Fe$_2$Ti)O$_4$ magnetite, (Fe$_{0.5}$Ti$_{0.5}$, Mg$_{0.5}$Ti$_{0.5}$,Fe$^{3+}$)O$_{1.5}$ ilmenite, (Fe,Ni,Si) metal, (CaAl,NaSi,KSi)AlSi$_2$O$_8$ feldspar, (Mg,Fe)Al$_2$Si$_{2.5}$O$_9$) cordierite, and (Fe,Ni)$_3$C carbide. They attributed minor differences in T$_C$ from Grossman (1972) to their use of internally consistent data from Helgeson et al. (1978). They appear to be the only workers prior to Yoneda and Grossman (1995) to include rigorous treatment of nonideal solid solutions. Saxena and Eriksson (1983) presented a complete phase diagram for solar composition vapor from 1153 to 1773 K and 1 to 10$^{-6}$ bar (log P vs. T).

Fegley and Palme (1985) extended calculations by Palme and Wlotzka (1976) to explain Mo and W depletions relative to other trace elements in some CAIs as due to CAI condensation at oxygen fugacities much higher than for a vapor of solar composition. However, Beckett (1986) showed by experiment that much lower oxygen fugacities are necessary to explain Ti$^{+3}$-bearing pyroxenes in igneous CAIs. Palme and Fegley (1990) calculated the oxygen enrichment necessary to form FeO-rich rims on olivines like those that occur in matrix and chondrule rims in the Allende (CV3) chondrite, at temperatures sufficient for Fe to diffuse into olivine. They discussed various mechanisms that might produce oxidizing conditions in the solar nebula.

Tsuchiyama and Kitamura (1995) calculated solid-gas equilibria in the system Fe-Mg-Si-O-C-H and presented simple ternary phase diagrams to illustrate potential crystallization paths under H-rich conditions. They stated that "the numbers of major gaseous species are the same as those of components in the concerned systems", therefore the calculations are very simple. They pointed out a 'thermal divide' above 500-700 K between O-poor and O-rich compositions and hypothesized that the redox states of ordinary versus enstatite chondrites might reflect different additions of $H_2O$ and $CH_4$-rich ices to source regions.

Sharp and Wasserburg (1995) adapted the SOLGASMIX program for the calculation of condensation in C-rich stellar envelopes,= in order to constrain the formation conditions of presolar graphite, TiC, SiC and AlN. Burrows and Sharp (1999), using 330 gas species and ~120 potential condensates, calculated condensation and potential rainout of solids from dwarf stars, with particular attention to the infrared spectrum of Gliese 229B and the newly discovered L dwarfs with effective T above 1500 K. They used NASA polynomials for thermodynamic data and did not consider the effect of gravitational settling of primary condensates.

Wood and Hashimoto (1993) developed a second-order method in their program PHEQ (Table 2), updated by Petaev and Wood (1998), following White et al. (1958). PHEQ treats 10 elements, 93 gas species, 148 condensed species, and assumes all solid solutions and melt to be ideal. Wood and Hashimoto (1993) investigated nebular systems fractionated based on volatility, considering four components: refractory dust (including Na and S), more volatile C-rich tar, ices, and H-rich gas. They found small stability fields for silicate melt at 1000x dust enrichment and $P^{tot} = 10^{-5}$ bar. The condensation behavior of FeS calculated using PHEQ differs markedly from results by other workers.

Yoneda and Grossman (1995) incorporated a nonideal model for $CaO-MgO-Al_2O_3-SiO_2$ (CMAS) liquid (Berman, 1983) to pioneer the investigation of liquid stability in systems enriched above solar composition by an ordinary chondrite-like composition dust similar to the refractory dust of Wood and Hashimoto (1993). Their code was based on the one used by Lattimer et al. (1978) and included solid solution models for metal, olivine, clinopyroxene, orthopyroxene, melilite and feldspar. Their results were consistent with many observations of igneous (melted) CAIs. Sylvester et al. (1990) had previously calculated the condensation of metals into refractory metal nuggets (a.k.a., Fremdlinge, El Goresy et al., 1978) during or prior to CAI formation.

Fedkin and Grossman (2006) adapted the code of Yoneda and Grossman (1995) to revisit nebular FeO condensation into olivine, arguing that diffusion presents a significant barrier to matching olivine compositions found in some chondrules in ordinary chondrites. Fedkin et al. (2010) used a revised version of the same code to revisit the condensation of reduced phases from calculated supernova shells and their mixtures, with careful attention to P-T conditions. They found it difficult to match the observed isotopic and chemical properties of TiC and metal grains found inside presolar graphite spherules formed in supernovae (Croat et al., 2003).

Campbell et al. (2001, 2002) built on earlier methods (Fegley & Palme, 1985; Palme & Wlotzka, 1976) to model metal condensation using $f(O_2)$-T relations calculated by Ebel and Grossman (2000) for solar compositions. They included gaseous oxides of V, Fe, Mo and W, and considered these and platinum-group element (PGE) Ru, Os, Rh, Ir, Pt, Pd and Ni gas/metal partitioning. To explore zoning in metal grains, two models were used: rapid cooling relative to metal condensation, for continued supersaturation of the gas in PGEs; and fractional condensation to enrich cores in PGEs, followed by diffusion to generate the zoning patterns observed in CB chondrite metal (Meibom et al., 2001). They concluded that a high-density environment was required to form zoned CB metal grains, consistent with a protoplanetary impact including a metal-rich body. Berg et al. (2009) built on this work to explore condensation of refractory metal nodules in chondrites.

Blander et al. (2004) built on arguments by Blander and Katz (1967) and calculated condensation with nucleation constraints in the system H, C, O, Si, S, Fe, Na, Ca, Al, Mg, Cr, Mn, Ti. They argued that constraints on the formation of high surface-energy solids stabilize silicate liquid droplets to >400 K below their liquidus and that solid or liquid Fe-Ni alloys are blocked entirely due to their high surface energy and tension, respectively. Blander et al. (2004) argued that these factors facilitated formation of FeO-rich olivine, even in a reduced system of solar composition. Grossman et al. (2012) tested these constraints using Equation 1 of Blander and Katz (1967). They showed that even if corundum and refractory metals like Os and W were supersaturated, the equation predicts that they would have nucleated homogeneously at T well above the equilibrium $T_C$ of Fe metal, providing sites for heterogeneous nucleation of both silicates and metal. Indeed, refractory metal nuggets (RMNs) are common in high-T condensates, found in CAIs, chondrules, and matrix (Daly et al., 2017a, 2017b), as well as in presolar graphite grains (Croat et al., 2013).

*VAPORS*

Ebel and Grossman (2000) considered the Berman (1983) CMAS liquid at high T, and the multi-component liquid model MELTS (Ghiorso & Sack, 1995) at T below the $T_C$ of olivine $(Mg,Fe)_2SiO_4$. This is necessary because the endmember species of the MELTS igneous rock crystallization model, developed for terrestrial silicate liquids, require substantial $SiO_2$ contents in the melt. The code treats 23 elements, ~500 gaseous species, and many condensed phases using internally consistent data compiled by Berman (1983, 1988) and Berman and Brown (1985), and the data used in the MELTS liquid calibration, whenever possible. The VAPORS code replaces the liquid reservoir of MELTS with a vapor, using c. 1995 MELTS algorithms to calculate a second-order approximation to the Gibbs energy surface for minimization, evaluate potentially stable phases, and minimize Gibbs energy of the entire system. Potential condensates include all the nonideal solid solutions calibrated for the MELTS software as well as silicate liquids (Ebel et al., 2000). VAPORS incorporated a new, nonideal solution model for Fe-Ni-Si-Cr-Co alloy, and correctly calculated (from Cp regressions) the Gibbs energies of formation of $C_2N_2$, $C_2H_2$, CN and HS gases incorrectly computed in the JANAF compilation (Chase, 1998; cf., Lodders, 1999, 2004).

Ebel and Grossman (2000; cf. Ebel, 2006, Plate 10) calculated silicate liquid stability in solar composition vapor and systems enriched in condensable elements by a CI chondrite-like dust (**Figure 3**). They found that conditions of dust enrichment >20x stabilize chondrule-like liquids and that dust enrichments of 100 to 1000x solar at $P^{tot} > 10^{-3}$ bar are required to stabilize the FeO-rich olivine found in type II chondrules. Ebel (2006) expanded these calculations to present a P-T phase diagram for vapor of solar composition from 1 to $10^{-8}$ bar $P^{tot}$, 1100 to 2000 K (their Plate 7), a T-C/O ratio diagram from 0.86 < C/O < 2.0 at $P^{tot} = 10^{-3}$ bar (their Plate 9), and a T-dust enrichment diagram for 1(solar) < C1 dust enrichment < 1000 at $P^{tot} = 10^{-3}$ bar (their Plate 10; see Figure 3).

Ebel (2000) varied metallicity to explore the correspondence between observed cold, dense interstellar cloud depletions and volatility-based depletion (cf., Field, 1974). Ebel and Grossman (2001) explored condensation in supernova shells depleted in molecular species due to Compton electron flux (Clayton et al., 1999), finding that TiC and SiC are not stable, but $SiO_2$ is stable, contrary to the evidence in presolar grains (Croat et al., 2003). Ebel and Grossman (2005) calculated condensation from expected vapor plume compositions for the Chicxulub (K-Pg) impact, finding that spinel compositions at the global K-Pg boundary are consistent with a thick carbonate target (i.e., Chicxulub) and not alternatives (e.g., granite, basalt).

Ebel and Alexander (2011) calculated condensation sequences for systems enriched in a reduced, interplanetary dust particle (IDP)-like dust, finding that minerals like CaS (oldhamite) and MgS (niningerite) become stable at high T, and silicates are FeO-free, as in enstatite chondrites and in Mercury's upper mantle (Ebel & Stewart, 2018). Ebel and Sack (2013) calculated the stability of the K-, Cl-bearing metal sulfide djerfisherite in these reduced conditions, finding that its petrological setting in EH3 chondrites is consistent with its calculated $T_C$ relative to chondrule liquids and associated minerals.

Ebel et al. (2012) calculated the thermochemical stability of low-iron, Mn-enriched "LIME" olivines, finding that conditions of their stability do not overlap with the T-P range for silicate liquid stability (high $P^{tot}$, high dust enrichment). This work relied on careful calibration of the solid solution behavior of Mn in olivine (Hirschmann, 1991; Hirschmann & Ghiorso, 1994). Constraints were placed on the conditions of formation of amoeboid olivine aggregates (AOAs), IDPs, and certain grains from comet Wild 2 returned by the Stardust mission. This work demonstrates the power of solid solution activity models in constraining formation conditions of natural samples.

Fedkin et al. (2012) integrated a closed-system evaporation model into a new version of the VAPORS code (this author has not seen this code), including provision for isotope fractionation. They concluded that chondrules formed by shock waves would record heavy isotope enrichments that are not observed. They therefore favored chondrule formation by impacts on ice-rich planetesimals. Fedkin and Grossman (2013) calculated evaporation of condensed droplets from their fully-melted (liquidus) T at assumed cooling rates, arguing that the enrichment of some chondrules in Na requires an impact origin (cf., Alexander et al., 2008). They calculated impact conditions for the formation of chondrules containing olivine of specific FeO contents with liquids of specific Na contents. Fedkin et al. (2015) investigated potential condensates from a variety of potential colliding bodies to explore the potential origin of chondrules in CB chondrites (Campbell et al., 2002). Fedkin and Grossman (2016) investigated condensation from plumes created by impacts of bodies of various (extreme) compositions and concluded that water-rich parent bodies are "excellent candidate sources of chondrule precursors".

The VAPORS code in its current state requires full 128-bit (32 digits of precision, "long double") precision to approximate the first and second partial derivatives of the Gibbs free energy surface of the speciated vapor phase with composition (Ebel et al., 2000). This makes the code difficult to implement on current "little-endian" processor architectures. Remarkable improvements have been made to the MELTS algorithms since 1995 (Gualda et al., 2012; Ghiorso, 2013) and are in development (Ghiorso, 2004). Work is underway to blend these efforts into a modern, consistently updated, open-source software tool (ThermoSNCC, Boyer, 2020; Boyer et al., 2019, 2020; ENKI, 2020).

*CONDOR*

The canonical 50% condensation temperatures ($T_{c50\%}$) for all the naturally occurring elements have been calculated using CONDOR (Lodders, 2003). This code was described by Lodders and Fegley (1993) with data sources in Fegley and Lodders (1994). It was reviewed by Fegley and Schaefer (2010). CONDOR has been applied to REE in C-rich systems (Lodders & Fegley, 1993), presolar SiC grain origin (Lodders & Fegley, 1995), condensation in C-rich stellar atmospheres (Lodders & Fegley, 1997a, 1997b), brown dwarf atmospheres (Fegley & Lodders, 1996), planetary atmospheres (Lodders & Fegley, 1994), and other problems (CONDOR, 2019). Precursors include METKON (Fegley & Palme, 1985; Kornacki & Fegley, 1986) and TOP20 (Barshay & Lewis, 1976, 1978). The code uses BNR methods (Smith & Missen, 1982), minimizing residuals in mass balance (error <$10^{-5}$) to solve gas-phase chemistry and gas-solid condensation simultaneously using iterative techniques. The code is written in PowerBasic, using "extended-precision" arithmetic (80 bits, or 18 digits of precision). As is true for most condensation codes, a weakness of this code is its lack of detailed thermodynamic activity models for complex, nonideal solid solution phases. The condensation of trace elements into pure solid hosts is accomplished by application of activity coefficients, when those have been established by experiments, as described by Kornacki and Fegley (1986). Strengths include coverage of all the elements and a very complete database for over 3600 solid, liquid and gaseous compounds including ions. CONDOR never included the errors present in the 3$^{rd}$- and 4$^{th}$-edition JANAF tables (e.g., for HS$_{(g)}$; Lodders, 2004b).

*Other Codes*

Efforts in chemical equilibrium modeling for cosmochemistry include those listed in Table 2. These codes have been applied to a large number of problems in cosmochemistry and planetary science. The NASA Glenn Research Center code "Chemical Equilibrium with Applications" (CEA) predates all of these efforts (CEA, 2019; Huff et al., 1951) but its primary focus has been problems associated with jet propulsion (Brinkley 1947; McBride & Gordon, 1996; White et al., 1958). Open-source development has ceased (C. Snyder, July 2019, personal communication), with a MatLab version available (U.S.-use only) in the NASA software catalog. The Cantera suite (Cantera, 2019) fills the CEA gap for application to problems in combustion, detonations, electrochemical energy conversion and storage, fuel cells, batteries, aqueous electrolyte solutions, plasmas, and thin film deposition. While one may pull algorithms or data from these efforts, they do not provide the necessary tools for dynamical cosmochemistry.

**Table 2:** Selected non-commercial codes developed for direct calculation of equilibrium solid (and liquid) condensation.

| code | developer | language | open | currency | n(Z) | n(op) | Reference |
|---|---|---|---|---|---|---|---|
| Top20 | J. Lewis | fortran | no | 1978 | 20 | | Barshay & Lewis (1976, 1978) |
| METKON | Palme et al. | fortran | no | 1985 | | | Fegley & Palme (1985) |
| CONDOR | K. Lodders | BASIC | no | ongoing | all | all | Lodders & Fegley (1993) |
| PHEQ | J. Wood | fortran | available | 1993 | 10 | 19 | Wood & Hashimoto (1993) |
| CWPI | M. Petaev | fortran | ** | 1998 | 19 | 19 | Petaev & Wood (1998) |
| ZONMET | M. Petaev | fortran | ** | 2003 | 19 | 19 | Petaev et al. (2003) |
| GRAINS | M. Petaev | fortran | no | ongoing | 19 | 19 | Petaev (2009) |
| CEA | C. Snyder | fortran | yes | c. 2004 | 50 | 20 | Gordon & McBride (1994); CEA (2019) |
| CWIN | S. Yoneda | fortran | no | 1995 | 23 | 23 | Yoneda & Grossman (1995) |
| VAPORS | D. Ebel | C | no | 2019 | all | 24 | Ebel & Grossman (2000); Ebel et al. (2000) |
| ThermoSNCC | G. Boyer | python, C | yes | ongoing | 10 | 10 | Boyer et al., 2019, 2020; ENKI |

*Currency is the most recent inferred year of active code development.

** n(Z) is the number of elements with data available.

*** n(op) is the operational maximum number of elements.

† Earlier versions of CWPI and ZONMET available from Petaev (2009) and Wood and Hashimoto (1993).

*Commercial Software*

The IVTANTHERMO for Windows code accesses data for 3200 substances formed by 96 chemical elements (Belov, 2015). Its calculation engine "EQUICALC" can handle only up to 700 substances, 60 pure phases, and a couple of condensed mixtures in a calculation including the gas phase. The combination has been used for calculations of chondrite and planetary outgassing (Schaefer &Fegley, 2007, 2010).

The FactSage software (closed source, downloadable) descends from ChemSage (Eriksson & Hack, 1990), which was a descendant of SOLGASMIX (Eriksson, 1971; Eriksson & Rosen, 1973; Saxena & Eriksson, 1983), both of which are heavily cited in the technical literature. It was linked with the comprehensive database of Barin (1991) and allowed users to choose solid solution models and parameters (Eriksson, 1975). SOLGASMIX formed the basis for several similar programs (e.g., Besmann, 1977), including that of Sharp and Huebner (1990) who applied it to the astrophysical problem of opacity but used the NASA polynomials for thermodynamic properties, which are valid only above 1000K and contain errors (Lodders, 2002, her Appendix). FactSage is advertised as "one of the largest thermodynamic software and database packages in the world", primarily applied in the metallurgical and ceramics industries (FactSage, 2020). Recent applications of FactSage for condensation include Pasek et al. (2005) and Pignatale et al. (2016).

The software name "HSC Chemistry" refers to enthalpy (H), entropy (S) and heat capacity ($C_P$). This proprietary software (now in version 10) has many of the features of FactSage (HSC Chemistry, 2020). It does not handle non-ideal solid solutions. Recent applications of HSC Chemistry for condensation include those of Pasek et al. (2005), Bond et al. (2010, HSC v. 5.1), and Elser et al. (2012).

**Condensation in Astrophysical Models**

Feedback between chemistry and dynamics is very limited or missing in recent applications of chemistry codes to astrophysical scenarios. The tools do not yet exist that would allow the results of chemistry to affect opacities, radiative transfer, settling or aerodynamic parameters that would then affect the next step(s) in dynamical calculations. Part of the difficulty is that commercial chemical codes do not 'talk to' dynamical codes.

Many astrophysicists have used commercially available software originally developed for the metallurgical industry. Earlier work linking nebular dynamics (e.g., diffusion, drift) to chemistry includes Cyr et al. (1999), who varied O in otherwise solar compositions, based on $H_2O$ depletions expected from calculations of water advection and diffusion in the inner nebula (Cyr et al., 1998). They then, separately, used the SOLGASMIX chemical equilibrium code (Table 2) to calculate condensation sequences for each $H_2O$ depletion. Bond et al. (2010) used HSC to model terrestrial planet formation at seven points in a model, evolving solar system by partitioning 16 elements between 33 solid species and 84 gaseous species (including 16 monatomic). Elser et al. (2012) obtained disk chemical gradients by applying the same HSC chemical approach (Bond et al., 2010) to a variety of disk models with different initial mass distributions.

Pasek et al. (2005) used a 2D steady-state α-disk solution to yield time - T - $P^{tot}$ histories in the midplane at each solar radius. They then computed condensation fronts. With a model for diffusive transport driven by condensation fronts in the inner disk, they then calculated condensation histories at progressively more oxygen-depleted inner radii. Pasek et al. (2005) used the SOLGASMIX code in parallel with HSC Chemistry to study the potential effects of water depletion on nebular S speciation, in the context of forming reduced rocks (e.g., enstatite chondrites) in a simple solar abundance protoplanetary disk model. They considered only ideal solid solutions and each code allowed different combinations of elements and species, while yielding "similar results" (Pasek et al., 2005) to those of Yoneda and Grossman (1995) at $P^{tot} = 10^{-6}$ bar, who included solid solution models for melilite and other phases.

Moriarty et al. (2014) used a disk model (Chambers, 2009) to calculate disk T, $P^{tot}$, and density over time. Earlier workers (e.g., Bond et al. 2010), had used HSC to calculate equilibrium planetesimal composition in disk regions. Moriarty et al. (2014) calculated chemistry at time and radius steps using HSC, removing gas and dust into planetesimals growing at a prescribed rate and decoupled from the gas. They also accounted for effects of radial gas movement (with perfectly coupled dust) on chemical inventory. Finally, their planetesimals were input into an *N*-body simulation to investigate redistribution of rock and ice during late-stage planet migration.

Pignatale et al. (2016) used disk models (D'Alessio et al., 1999) to prescribe T and $P^{tot}$ in a 2D disk, then ran condensation using FactSage software from high T to those conditions. They then, separately, applied dust settling and radial migration models (Liffman & Toscano, 2000) to calculate 2D chemical distributions of material and the extent of a non-ionized "dead zone" (Lesur et al., 2014) after estimated equilibration times.

Li et al. (2019) used an evolving disk model (Cassen, 1996) and the "GRAINS" code (Petaev, 2009) to calculate condensed solid abundances. Their final distribution of refractory and more volatile elements in radial zones of the disk depends upon the decoupling timescale of solids from gas.

## Outlook

Bridging the gap between cosmochemical calculations and dynamical models of protoplanetary disks and dust evolution is both a difficult challenge and one of the greatest opportunities facing the study of planet formation and the origin of our Solar System. Given the complexity of cosmochemical samples like chondritic meteorites, such a bridge will by necessity be largely numerical. Commercial software lacks the flexibility to efficiently integrate with dynamical modeling, but modern, consistently updated, open-source tools for condensation calculations are in development (Boyer et al., 2019, 2020b).

Why is it important to integrate chemical calculations with dynamical models? From the perspective of numerical simulations, condensation and evaporation calculations will affect dust porosity and hence aerodynamics, opacity and how much the dust promotes recombination and thus reduces the ionization fraction. The ability to determine when the condensed assemblage supports a stable liquid phase or phases is therefore critical: collisional coagulation leads to porous dust aggregates with large surface-area to mass ratios, while melting leads to coalescence and low surface area to mass ratios. Also important are the generation and implications of water ice cold traps for the thermal processing of solids, and the effect of magnetized grain interactions on the vertical distribution of siderophile-rich grains. All these feedbacks make condensation and evaporation chemistry critical to modeling of dynamical processes in planet formation.

## Acknowledgments

This article benefited greatly from comments by two reviewers, and discussion and suggestions by Dr. Katharina Lodders. Asna Ansari contributed to the section on associated solutions. The material is based upon work partially supported by the National Aeronautics and Space Administration under Grant No. NNX16AD37G issued through the Emerging Worlds program (DE). The research made use of NASA's Astrophysics Data System.

## Further Reading

Grossman, L. (1975). The most primitive objects in the solar system. *Scientific American*, *232*, 30–39.

[https://doi.org/10.1088/1742-6596/946/1/012120]. *Journal of Physics: Conference Series, 946*.

Berg, T., Maul, J., Schoönhense, G., Marosits, E., Hoppe, P., Ott, U., & Palme H. (2009). Direct evidence for condensation in the early solar system and implications for nebular cooling rates. *The Astrophysical Journal Letters, 702*, L172–L176.

Berman, R. G. (1983). A thermodynamic model for multicomponent melts, with application to the system $CaO-MgO-Al_2O_3-SiO_2$. Doctoral thesis, University of British Columbia, Vancouver, 178 pp.

Berman, R. G. (1988). Internally consistent thermodynamic data for minerals in the system $Na_2O-K_2O-CaO-MgO-FeO-Fe_2O_3-Al_2O_3-SiO_2-TiO_2-H_2O-CO_2$. *Journal of Petrology, 29*, 445-522.

Berman, R. G., & Brown, T. (1985). Heat capacity of minerals in the system $Na_2O-K_2O-CaO-MgO-FeO-Fe_2O_3-Al_2O_3-SiO_2-TiO_2-H_2O-CO_2$: representation, estimation, and high-temperature extrapolation. *Contributions to Mineralogy & Petrology, 89*, 168-183.

Bernatowicz, T., Fraundorf, G., Ming, T., Anders, E., Wopenka, B., Zinner, E., & Fraundorf, P. (1987). Evidence for interstellar SiC in the Murray carbonaceous meteorite. *Nature, 330*, 728-730.

Bernatowicz, T. J., Cowsik, R., Gibbons, P. C., Lodders, K., Fegley, B. Jr., Amari, S., & Lewis, R. S. (1996). Constraints on stellar grain formation from presolar graphite in the Murchison meteorite, *Astrophysical Journal, 472,* 760-782.

Bernatowicz, T., Akande, W. O., Croat, T. K., & Cowsik, R. (2005). Constraints on grain formation around carbon stars from laboratory study of presolar graphite. *Astrophysical Journal, 631*, 988-1000.

Besmann, T., M. (1977). SOLGASMIX-PV: A computer program to calculate equilibrium relationships in complex chemical systems. Oak Ridge National Lab Report TM-5775.

Black, D. C., & Pepin, R. O. (1969). Trapped neon in meteorites - II. *Earth and Planetary Science Letters, 6*, 395-405.

Blander, M., & Katz, J. L. (1967). Condensation of primordial dust. *Geochimica Cosmochimica Acta*, *31,* 1025-1034.

Blander, M., Pelton, A. D., Jung, I-H., & Weber, R. (2004). Non-equilibrium concepts lead to a unified explanation of the formation of chondrules and chondrites. *Meteoritics & Planetary Science, 39*, 1897-1910.

Bolser, D., Zega, T. J., Asaduzzaman, A., Bringuier, S., Simon, S. B., Grossman, L., Thompson, M. S., & Domanik, K. J. (2016). Microstructural analysis of Wark-Lovering rims in the Allende and Axtell CV3 chondrites: Implications for high-temperature nebular processes. *Meteoritics & Planetary Science, 51*, 743-756.

Bond, J. C., Lauretta, D. S., & O'Brien, D. P. (2010). Making the Earth: Combining dynamics and chemistry in the Solar System. *Icarus, 205*, 321-337.

Boyer, G. (2020). ThermoSNCC code repository [https://gitlab.com/gmboyer/thermosncc] (accessed 12 October 2020)

Boyer, G., Unterborn, C., Ebel, D. S., Ghiorso, M. S., & Desch, S. (2019). ThermoSNCC: a free tool for modeling condensation sequences. *Eos Trans. AGU* 2019, Abstract#601733
24

**Denton S. Ebel**